\documentclass[letter, scriptaddress, twocolumn,  prd, showpacs, showkeys]{revtex4-1}

	\usepackage{amsmath}
    \usepackage{mathrsfs}
	\usepackage{amssymb}
	\usepackage{graphicx} 
	\usepackage{makeidx}
	\usepackage{amsfonts}
	\usepackage[ansinew]{inputenc}
	\usepackage[usenames, dvipsnames]{pstricks}
	\usepackage{epsfig}
	\usepackage{pst-grad} % For gradients
	\usepackage{pst-plot} % For axes
	\usepackage[colorlinks, hyperindex]{hyperref}
	\usepackage{float}
	\usepackage{lipsum}
	\usepackage{makecell}
	\usepackage[caption=false]{subfig}
\begin{document}
\title{{\Large\bf Generalized Brans-Dicke Theory: A Dynamical Systems Analysis}}

\author{Nandan Roy } 
\email{nandan@fisica.ugto.mx}
\affiliation{Harish-Chandra Research Institute,\\ Chhatnag Road, Jhunsi, Allahabad -211019, India.}
\altaffiliation[Present address:]{ Departamento de Fisica, DCI Campus Leon Universidad de Guanajuato, 37150 Leon, Guanajuato, Mexico.}

\author{Narayan Banerjee} 
\email{narayan@iiserkol.ac.in}
\affiliation{Department of Physical Sciences,~~\\Indian Institute of Science Education and Research Kolkata,~~\\Mohanpur Campus, West Bengal 741246, India.}

%\begin{center}
%{\Large\bf Generalized Brans-Dicke Theory: A Dynamical Systems Analysis}
%\\[15mm]
%Nandan Roy \footnote{E-mail: nandanroy@hri.res.in} and
%Narayan Banerjee \footnote{E-mail: narayan@iiserkol.ac.in}
%
%~~~~~
%
%{\em $~^{1}$Harish-Chandra Research Institute, \\ Chhatnag Road, Jhunsi, Allahabad - 211019, India. \\ ~~ \\ $~^{2}$Department of Physical Sciences,~~\\Indian Institute of Science Education and Research Kolkata,~~\\Mohanpur Campus, West Bengal 741246, India.}\\[15mm]
%\end{center}

\begin{abstract}

The stability criteria for the generalized Brans-Dicke cosmology in a spatially flat, homogeneous and isotropic cosmological model is discussed in the presence of a perfect fluid. The generalization comes through the channel that the Brans-Dicke coupling parameter $\omega$ is allowed to be a function of the scalar field $\phi$. This generalization can lead to a host of scalar-tensor theories of gravity for various choices of $\omega = \omega (\phi)$. A very interesting general result  has been found. Excepting for the case of a radiation distribution as the choice of the fluid, all other solutions find a natural habitat in the corresponding solutions in general relativity in an infinite $\omega$ limit. For the radiation distribution, the dependence of stability on $\omega$ is a bit obscure. If a scalar potential, function of the Brans-Dicke scalar field, is added to the action, the requirement of an infinite $\omega$ for stability is relaxed for a matter distribution with a non-zero trace whereas it becomes a possibility for a radiation distribution.
\end{abstract}



\maketitle
%PACS: 98.80.-k; 95.36.+x

\section{Introduction:}

Brans-Dicke (BD) theory\cite{brans}, a failed attempt to incorporate Mach's principle in a relativistic theory of gravity, is a simple scalar tensor extension of general relativity (GR) where a dynamical scalar field is nonminimally coupled to the curvature. This makes the Newtonian constant $G$ inversely proportional to the scalar field $\phi$ and hence a function of the coordinates. A dimensionless parameter $\omega$, called the Brans-Dicke coupling parameter, determines the deviation of the results obtained in this theory under weak field approximation from that in general relativity under similar approximation. The lower the value of $\omega$, the more different are the corresponding results. It is quite well known that general relativity does extremely well in explaining local astronomical tests, and the value of $\omega$ required such that BD theory is consistent with such observations are too high ($\omega > 500$) making BD theory practically indistinguishable from GR in the weak field limit\cite{will}. \\

An order of magnitude estimate showed that not only in the weak field approximation, in the limit $\omega \rightarrow \infty$,  $\phi$ becomes a constant and its value behaves as $\frac{1}{\omega}$ which makes the set of field equations, in its full nonlinear generality, reduce to the corresponding GR equations\cite{wein}. This was believed to be a great advantage of BD theory as it gives the good old GR in some limit. This had a jolt, when it was clearly shown that this infinite $\omega$ limit has only a limited application and fails in the case where the matter content has a traceless stress-energy tensor (such as radiation, Maxwell field etc.)\cite{soma, valerio1}. \\

In spite of all these, BD theory enjoys a periodic resurgence for various reasons, particularly in cosmology. For instance, BD theory played a crucial role in suggesting an extended inflation\cite{johri, la} which proved so useful in eradicating the graceful exit problem of standard inflationary models. In more recent years, BD theory has been used to create a perfect ambiance for a late time acceleration for the universe\cite{many}. The remarkable feature is that BD theory, in its own right, can generate an accelerated expansion, without any exotic field, only by a suitable choice of value for the parameter $\omega$\cite{nbdp}. \\

In order to incorporate the various requirements for the value of the parameter $\omega$ for different situations, a generalization of BD theory, where $\omega$ is taken to be a function of the scalar field $\phi$ was also proposed quite a long time back\cite{berg, wago, nord}. It was found that many other nonminimally coupled scalar tensor theories of gravity, suggested so as to describe different requirements, can be written as a special case of the generalized BD theory by a suitable choice of $\omega = \omega (\phi)$. For example, one recovers Barker's scalar tensor theory\cite{barker} for $2\omega + 3 = \frac{1}{\phi - 1}$ and Schwinger's theory \cite{schw} for $2\omega + 3 = \frac{1}{\alpha \phi}$ where $\alpha$ is a constant. For a brief collection of such examples and the physical motivation of various choices of $\omega$ as a function of $\phi$, we refer to the work by Van den Bergh\cite{vanden}. It deserves mention that even for a varying $\omega$, the structure of Einstein field equations remain the same as they come from the variation of the relevant action with respect to the metric tensor and not the scalar field. Thus the conclusions regarding the infinite $\omega$ limit of the generalized theory will be the same as that for the BD theory with a constant $\omega$. \\

As BD theory has application mostly in cosmology, various aspects of the cosmological solutions have been fairly well-studied. The stability of such solutions for a standard isotropic and homogeneous cosmology through a phase space analysis has been studied extensively by Kolitch and Eardley\cite{kolit}. A very interesting result obtained in the work is that models with non flat spatial curvature with a vacuum energy leading to an inflationary expansion in the past do not approach the corresponding spatially flat solutions in late time. Holden and Wands\cite{wands} extended this work in the sense that their analysis takes care of all Friedmann-Robertson-Walker solutions in Brans-Dicke theory. The latter investigation had been worked out in the conformally transformed version of BD theory, where the nonminimal coupling is apparently broken by paying the price of the validity of the equivalence principle. A very interesting result of this work is that all the stable solutions, lasting for an infinite future, the BD parameter $\omega$ has values less than $-\frac{4}{3}$. This result is crucial, as the requirement for an accelerating universe in BD theory, without any exotic matter, is a negative $\omega$ with a low value\cite{nbdp}. Santos and Gregory\cite{ruth} used the method of dynamical systems analysis for BD theory where the models are endowed with a potential $V=V(\phi)$. This kind of extended BD theory finds application in finding a framework for the accelerated expansion of the universe\cite{bertolami}.\\

The present work deals with a generalized version of BD theory where the parameter $\omega$ is a function of the scalar field in the framework given by Nordtvedt\cite{nord}. The motivation is to see the stability of the solutions for a very general choice of $\omega (\phi)$. As the matter content, a perfect fluid with a barotropic equation of state is considered. The work closest to the present work in the literature is the one by Faraoni\cite{valerio2}. Faraoni's work also starts with a variable $\omega$ in the same framework, and then specializes to either BD theory itself with a potential or a scalar tensor theory where $\omega = \frac{G\phi}{4\xi(1-G\phi)}$, where $G$ is the Newtonian constant of gravity and $\xi$ is a constant. \\

The present work also includes a generalization in the form of the presence of a potential $V =V(\phi)$. The stability criteria is found to differ as expected. Mimoso and Nunes had worked on such a generalization of BD theory with either a radiation or a cosmological constant as the matter content. They worked in a conformally transformed version of the theory where the action looks like that of a minimally coupled theory, and arrived at the result that GR is an attractor of the BD theory\cite{mimoso}. The problem with the conformally transformed version is that the principle of equivalence is no longer valid as the rest mass of the test particles becomes a function of the BD scalar field. \\

In section II, we present the dynamical systems analysis where no additional potential is used and an additional potential is considered in section III. The fourth section includes a discussion of the results obtained.

\section{Dynamical System without potential}
The action of Brans Dicke(BD) theory in Jordan frame with a varying coupling constant is of the following form

\begin{equation} \label{action}
\mathcal{S} = \frac{1}{16 \pi} \int d^4 x \sqrt{- g}[ \phi R - \frac{\omega(\phi)}{\phi} g^{ab} \nabla_a \phi \nabla_b \phi] + \mathcal{S}^{(m)},
\end{equation} 

where $\mathcal{S}^{m} = \int d^4 x \sqrt{- g} \mathcal{L}_m$, describes the matter content, $\phi$ is Brans Dicke scalar field, which is minimally coupled to matter Lagrangian $\mathcal{L}_m$, but nonminmally coupled to Ricci scalar and $\omega(\phi)$ is a function of the BD scalar field. 

By varying the action with respect to the metric components, we obtain the field equations as

\begin{widetext}
\begin{equation} \label{field}
R_{ab} - \frac{1}{2} g_{ab} R = \frac{8 \pi}{\phi} T_{ab} ^{(m)} + \frac{\omega(\phi)}{\phi^2} (\nabla_a \phi \nabla_b \phi - \frac{1}{2} g_{ab} \nabla^c \phi \nabla_c \phi) + \frac{1}{\phi} (\nabla_a \nabla_b \phi - g_{ab} \Box \phi).
\end{equation}
\end{widetext}

Variation of the action with respect to $\phi$ gives us the Klein-Gordon equation for the scalar field as
\begin{equation} \label{phi}
\Box \phi = - \frac{\phi}{2 \omega} R - \frac{1}{2} (\nabla^c \phi \nabla_c \phi) (\frac{1}{\omega} \frac{d\omega}{d \phi} - \frac{1}{\phi}).
\end{equation}

Trace of the equation(\ref{field}) gives the expression for Ricci Scalar as
\begin{equation} \label{tr}
R = - \frac{8 \pi}{\phi} T^{(m)} + \frac{\omega}{\phi^2} \nabla^c \phi \nabla_c \phi + 3 \frac{\Box \phi}{\phi}.
\end{equation}

Substituting (\ref{tr}) in equation (\ref{phi}) and eliminating $R$, one obtains
\begin{equation}
\Box \phi = \frac{1}{2 \omega + 3} (8 \pi T^{(m)} - \frac{d\omega}{d \phi} \nabla^c \phi \nabla_c \phi).
\end{equation}

%Conservation equation,
%
%\begin{equation}
%\nabla^b T^{(m)} _{ab} = 0
%\end{equation}

We consider the universe to be described by the spatially flat FRW metric

\begin{equation}
ds^2 = - dt^2 + a(t)^2 [dr^2 +r^2 (d \theta^2 + \sin^2 \theta d\phi^2)].
\end{equation}

We also consider the universe to be filled with a perfect fluid with an equation of state $p= (\gamma -1) \rho$, where $p$ and $\rho$ denote the pressure and the density of the fluid respectively and $\gamma$ is a constant ($1 \leq \gamma \leq 2$). The field equations(\ref{field}) then can be written as
\begin{equation} \label{frid1}
3 H^2 = \frac{8 \pi}{\phi} \rho_{m} + \frac{\omega}{2} \frac{\dot{\phi}^2}{\phi^2} - 3 H \frac{\dot{\phi}}{\phi},
\end{equation}

\begin{equation} \label{frid2}
\dot{H} = - \frac{\omega}{2} (\frac{\dot{\phi}}{\phi})^2 - \frac{8 \pi}{\phi} \rho_{m} (\frac{2 + \gamma \omega}{3 + 2 \omega}) + 2 H \frac{\dot{\phi}}{\phi} + \frac{1}{2 (2 \omega + 3) \phi} \frac{d \omega}{d \phi} \dot{\phi}^2,
\end{equation}

where $H$ is the Hubble parameter $(H = \frac{\dot{a}}{a})$ and an overhead dot represents differentiation w.r.t the cosmic time t. The Klein Gordon equation for the BD Scalar field reduces to

\begin{equation} \label{wave}
\ddot{\phi} + 3 H \dot{\phi} = \frac{8 \pi}{2 \omega + 3} (4 - 3 \gamma) \rho_{m} - \frac{\dot{\phi}^2}{2 \omega + 3} \frac{d \omega}{d\phi}.
\end{equation}

Equation (\ref{frid1}) can be written in a dimensionless form
 
\begin{equation}
1 = \frac{8 \pi \rho_m}{3 H^2 \phi} + \frac{\omega}{6} \frac{\dot{\phi}^2}{H^2 \phi^2} - \frac{\dot{\phi}}{H \phi} .
\end{equation}
 
We introduce a new set of dimensionless variables, $x = \frac{\dot{\phi}}{H \phi}, y = \frac{1}{2 \omega + 3}, \lambda = \phi \frac{d \omega}{d \phi}$ and $N = \ln(\frac{a}{a_{0}})$, where $a_{0}$ is the present value of the scale factor which subsequently will be taken to be unity. With these variables, system of equations reduces to the following autonomous set of equations,

\begin{widetext}

\begin{equation} \label{x}
x^{\prime} = \frac{3}{2} (1 - \frac{1}{12 y} x^2 (1 - 3y) + x) [-(2 - \gamma) x + (2 + x) (4 - 3 \gamma) y] - \lambda x^2 y - \frac{1}{2} \lambda x^3 y,
\end{equation}

\begin{equation} \label{y}
y^{\prime} = - 2 \lambda x y^2,
\end{equation}

\begin{equation} \label{lambda}
\lambda^{\prime} = ({\Gamma}+\lambda) x.
\end{equation}

\end{widetext}

Here, $\Gamma = \phi^2 \frac{d^2 \omega}{d \phi^2}$ and a prime is the derivative with respect to $N = \ln a$.

For the qualitative analysis of the system with any functional form of $\omega (\phi)$, we classify our system into two classes. If (i) ${\Gamma} + \lambda = 0$ which leads to $\omega (\phi) = \ln (B \phi^A)$ where $A, B$ are constants  and if (ii) ${\Gamma} + \lambda \neq 0$, $\omega (\phi)$ is any function of $\phi$ except the functional form $\omega (\phi) = \ln (B \phi^A)$. In the case (i), $\lambda$ is a constant and the problem essentially becomes two-dimensional. The second case remains a three dimensional problem. These two classes are again classified into two more sub classes in the parameter space, $\gamma \neq \frac{4}{3}$ and $\gamma = \frac{4}{3}$ because the fixed points are qualitatively different for these two cases. The second kind with $\gamma = \frac{4}{3}$ indicates a radiation distribution and the first kind with $\gamma \neq \frac{4}{3}$ signifies any other kind of barotropic fluid. \\

\subsection{ Class I :${\Gamma} + \lambda = 0$: a 2 dimensional problem}
\subsection*{ i)~ $\gamma \neq \frac{4}{3}$}

In this case we have only one fixed point, namely $x = 0, y \rightarrow 0$. From the definition of $y$, $y \rightarrow 0$ implies $\omega \rightarrow \infty$. This physically means that the system has a fixed point at a very large value of $\omega$. One can also choose a different variable transformation where the variable is proportional  to $\omega$ and the fixed point at $\omega \rightarrow \infty$ can be found out directly. The reason behind our choice of $y = \frac{1}{2 \omega + 3}$ is the simple fact that zeroes are easier to handle than infinities. This kind of transformation of a variable has previously been utilized By Ng, Nunes and Rosati\cite{ng} in the context of a scalar field. It deserves mention that  instead of $x =0, y \rightarrow 0$ we can also write the fixed point as $x =0, y = 0$  although we have $\frac{1}{y}$ in the expression for $x^{\prime}$, it poses no real threat as the term actually has $\frac{x^{2}}{y}$, and if both $x$ and $y$ approach zero at the same rate, the numerator goes to zero much faster. \\

 The choice $y \rightarrow 0$ is made, because it simplifies stability analysis. There is a special class of non hyperbolic fixed points, called a normally hyperbolic fixed point which is basically a set of  non isolated fixed points and there is always a zero eigenvalue associated with each point \cite{coley}. Stability of normally hyperbolic fixed points can be found out easily from the sign of the remaining eigenvalues. If $y = 0 $, the fixed point is an isolated fixed point whereas for $y \rightarrow 0$, it is a non isolated fixed point. Throughout the whole work we have encountered some fixed points which have one zero eigenvalue. If we choose $y \rightarrow 0$, these fixed points are normally hyperbolic and the stability can be investigated without much of difficulties. The phase plot we have drawn for these fixed points also support our analytical finding considering $y \rightarrow 0$. \\

To check the stability of the fixed point we have to find the Jacobian matrix of the system at the fixed point.  The Jacobian matrix looks like,
 \[
J=
  \begin{bmatrix}
    -\frac{3}{2} (2 - \gamma)  &  3(4- 3 \gamma) \\
    0 & 0
  \end{bmatrix}
\]

The eigenvalues of the Jacobian matrix at this fixed point are $m_1 = -\frac{3}{2} (2 - \gamma), m_2 = 0$.  This is a non-isolated fixed point and it has one zero eigenvalue at each point, so the fixed point is a normally hyperbolic fixed point. The stability of a normally hyperbolic fixed point is determined from the sign of the remaining eigenvalues. In this case the fixed point is stable for $\gamma < 2$. For a fluid, the upper limit of $\gamma$ is $2$. So we indeed find that the extended BD theory, with an ideal fluid and a particular choice of $\omega$ as $\omega (\phi) = \ln (B \phi^A)$, the stable model is for $x=0$ and $ y \rightarrow 0$. The translation in terms of more physical quantities, the stability requires $\phi \rightarrow constant,~ \omega \rightarrow \infty$.

\subsection*{$ ii) ~\gamma = \frac{4}{3}$}
For $\gamma = \frac{4}{3}$, the fluid corresponds to a radiation distribution. The fixed point of the system is $x=0$ and $y$ is undetermined. The Jacobian matrix of the system at this fixed point is 
\[
J=
  \begin{bmatrix}
    -1  &  0 \\
    -2 \lambda y^2 & 0
  \end{bmatrix}
\]

The eigenvalues of the system at the fixed point are $m_1 = -1, m_2 = 0$. This is also a normally hyperbolic fixed point, so the fixed point is always stable. The physical behaviour of the system at this fixed point is $\phi \rightarrow constant$ and $\omega$ undetermined. \\

The clear indication is that for a radiation distribution, the stability of the Brans-Dicke solutions does not depend on the value of the parameter $\omega$ whereas for other varieties of fluid, for which the trace of the energy momentum is nonzero, the stability requires an infinite value of $\omega$. 

Fig 1 shows the phase plots of the system for both $\gamma \neq 4/3$ and $\gamma = 4/3$ cases. These plots strongly support our analytical findings.

\begin{figure}[h]

\begin{center}

\subfloat[Phase plot of the system for $\gamma \neq 4/3$ shows ($x=0, y \rightarrow 0$) is a stable fixed point.]{%
  \includegraphics[clip,width=0.6\columnwidth]{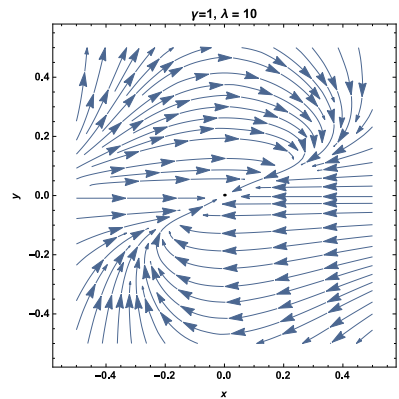}%
}

\subfloat[Phase plot of the system for $\gamma = 4/3$ shows ($x=0, y =$ undetermined ) is a stable fixed point.]{%
  \includegraphics[clip,width=0.6\columnwidth]{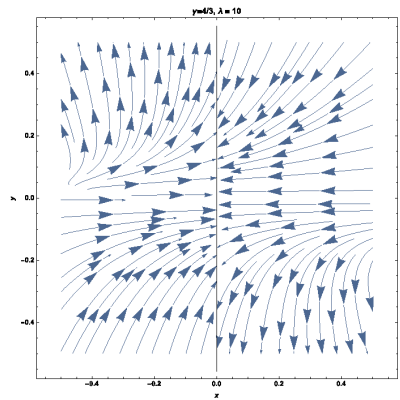}%
}

\caption{Phase plots of the class I system}
\end{center}
\end{figure}

\subsection{Class II: ${\Gamma} + \lambda \neq 0$: a 3 dimensional problem}

In this class ${\Gamma} + \lambda \neq 0$, so $\lambda$ is not a constant to start with and the problem remains a 3 dimensional one. Here $\omega$ is any function of $\phi$ excluding the functional form $\omega (\phi) = \ln (B \phi^A)$.  

\subsection*{$i) ~\gamma \neq \frac{4}{3}$}

In this case we have only one fixed point $x = 0, y \rightarrow 0, \lambda$.  The Jacobian matrix of the fixed point is given below
\[
J=
  \begin{bmatrix}
    -\frac{3}{2} (2 - \gamma)  &  3(4- 3 \gamma) & 0 \\
    0 & 0 & 0 \\
    \Gamma + \lambda & 0 & 0 \\
  \end{bmatrix}
\]

The eigenvalues of the fixed point are ($m_1 = -\frac{3}{2} (2 - \gamma), m_2 = 0, m_3 = 0 $ ). This is a nonhyperbolic fixed point. We can not use linear stability analysis in this case. So we resort to a different strategy to find the stability. We perturb the system from the fixed point by a small amount and find the evolution of the perturbations numerically. If the system comes back to the fixed point following the perturbation then the system is stable and if the perturbation grows so that the system moves away from the fixed point, then the system is unstable. This is basically the phase plot of the system near this fixed point. It is very difficult to draw conclusions from the 3D phase plot. So the projections of the perturbations on $x, y $ and $\lambda$ axis are considered separately. Figure 2 shows the results. Recently this technique is used in\cite{what}. As $N \rightarrow \infty$ the system comes back to $x =0, y \rightarrow 0$ but projection of perturbations increase monotonically along $\lambda$ axis. Our fixed point is $x =0, y \rightarrow0$, $\lambda$ arbitrary and after perturbation the system comes back to the same fixed point as $N \rightarrow \infty$. Hence we conclude that the fixed point is a stable fixed point. The physical state of the system at this fixed point is $\phi \rightarrow$ constant and $\omega \rightarrow \infty$. \\

\begin{figure*}[]
\begin{center}

\subfloat[Projection of perturbations along $x$ axis.]{%
  \includegraphics[clip,width=0.65\columnwidth]{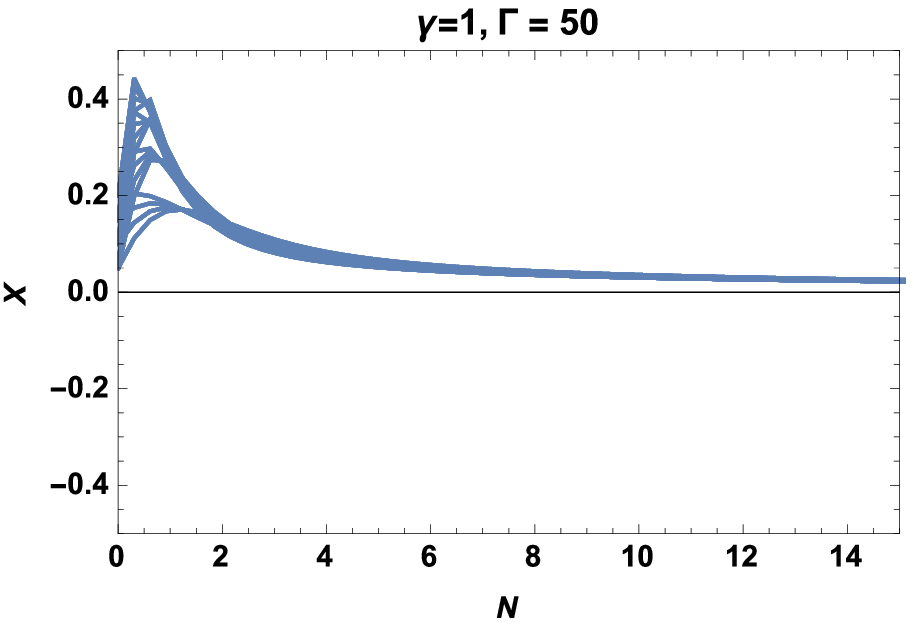}%
}
\subfloat[Projection of perturbations along $y$ axis.]{%
  \includegraphics[clip,width=0.65\columnwidth]{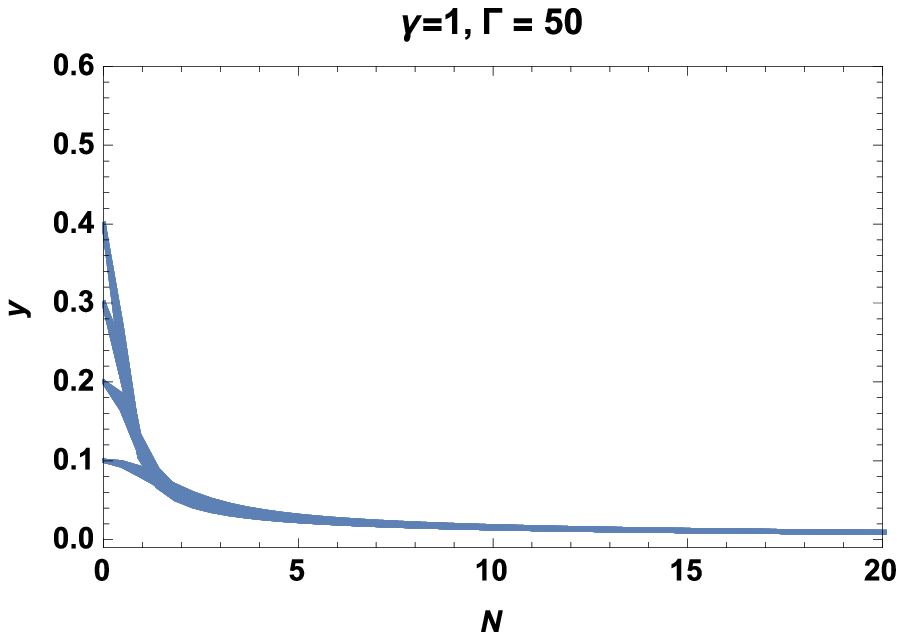}%
}
\subfloat[Projection of perturbations along $\lambda$ axis.]{%
  \includegraphics[clip,width=0.65\columnwidth]{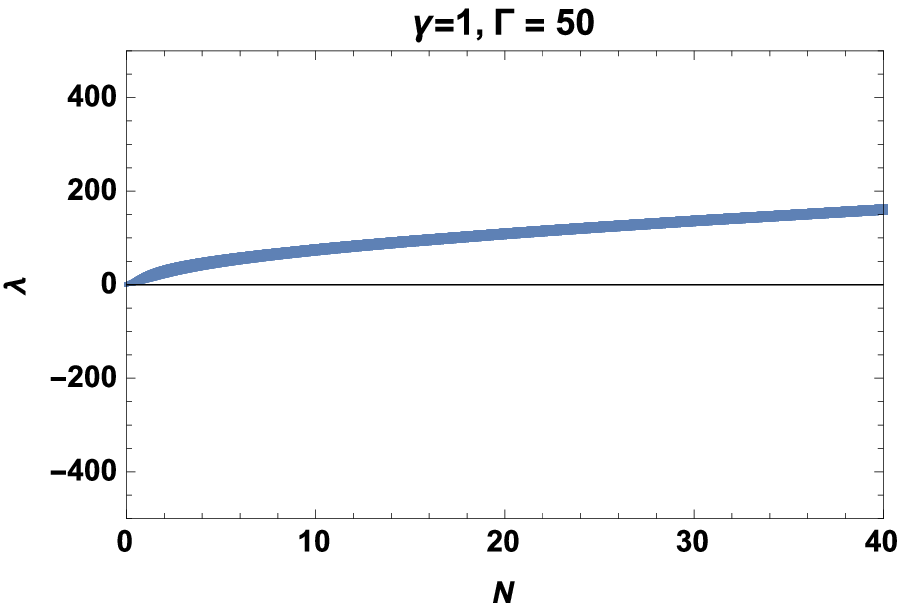}%
}

\caption{Projection of perturbations along $x,y , \lambda$ axis for Class II, without potential and $\gamma \neq \frac{4}{3}$.}

\end{center}

\end{figure*}

\subsection*{$ii)~ \gamma = \frac{4}{3}$}

In this case the fixed point  of the system is $x=0, y = $ undetermined, $ \lambda = $ undetermined. Eigenvalues are $m_1 = - \frac{3}{2} (6 - \gamma), m_2 =0, m_3 =0$. It is also nonhyperbolic in nature. So we have to explore the stability numerically as discussed before. Figure 3 shows the plots of perturbations along $x,y, \lambda$ axis. Projection of perturbation along $y$ and $\lambda$ axis evolve to some constant values. Perturbations about $x$ all converges to $x=0$, i.e., the fixed point value. Hence we can conclude that this fixed point is a stable fixed point. \\

Like the previous case where ${\Gamma} + \lambda = 0$, here also the requirement for stability is the same, an infinite $\omega$ for any matter distribution with a nonzero trace and no particular range of $\omega$ for radiation for which $T=T^{\mu}_{\mu}=0$. \\

It looks all a bit surprising why the requirement of the value of $\omega$ is so different for a radiation distribution and any other kind of fluid for which $T \neq 0$. While it does not have any imprint on the stability for the radiation case ($T=0$), a large value of $\omega$ plays a crucial role for other kinds of fluids. The clue is there in the Klein-Gordon equation (\ref{wave}) itself. One needs to have a trivial $\phi$, i.e., $\phi$= constant so as to generate the corresponding equations in General Relativity. Let us pick up the example of a constant $\omega$ which corresponds to the original Brans-Dicke theory. The first term in the right hand side of equation (\ref{wave}) is zero for $\gamma = \frac{4}{3}$, and the equation immediately yield a first integral as $\dot{\phi} a^{3} = \frac{\alpha}{(2\omega + 3)^{\frac{1}{2}}}$. So $\phi$ can be a constant if the constant of integration $\alpha$ is zero or $\omega\rightarrow \infty$. Thus an infinite $\omega$ is not a unique requirement. For any other distribution of matter, $\gamma \neq \frac{4}{3}$, and it is easy to see from equation (\ref{wave}) that $\omega\rightarrow \infty$ is an utmost requirement for having a constant $\phi$. This is true for any arbitrary functional dependence of $\omega$ on $\phi$. Fay studied a vacuum anisotropic model in a generalized BD theory and found that the stable solutions find their habitat either in string theory or in GR\cite{fay}. But this does not warrant an infinite value of $\omega$. For a vacuum, the energy-momentum tensor is indeed trace less, so the results obtained by Fay is consistent with the present analysis.

\begin{figure*}[]
\begin{center}

\subfloat[Projection of perturbations along $x$ axis.]{%
  \includegraphics[clip,width=0.65\columnwidth]{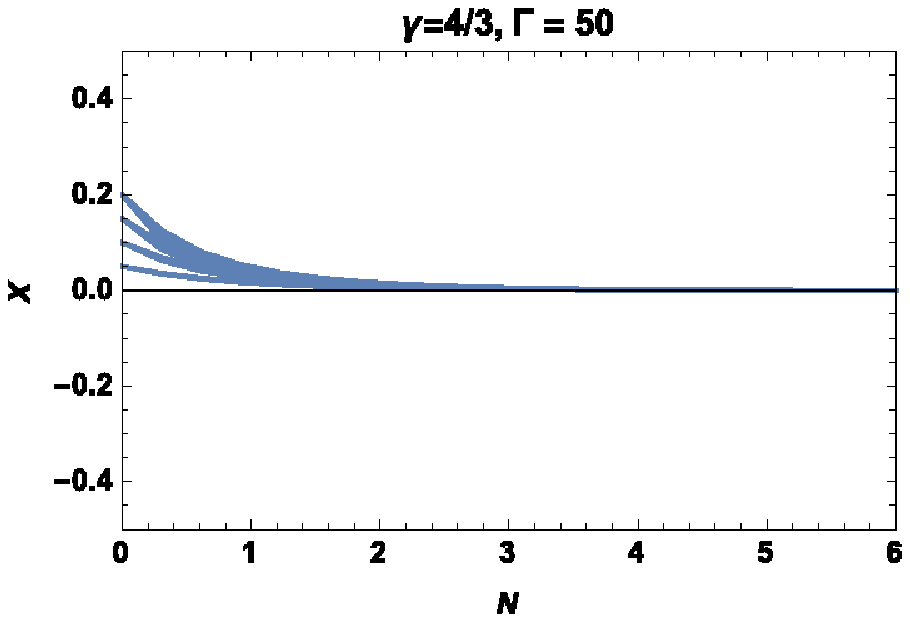}%
} 
\subfloat[Projection of perturbations along $y$ axis.]{%
  \includegraphics[clip,width=0.65\columnwidth]{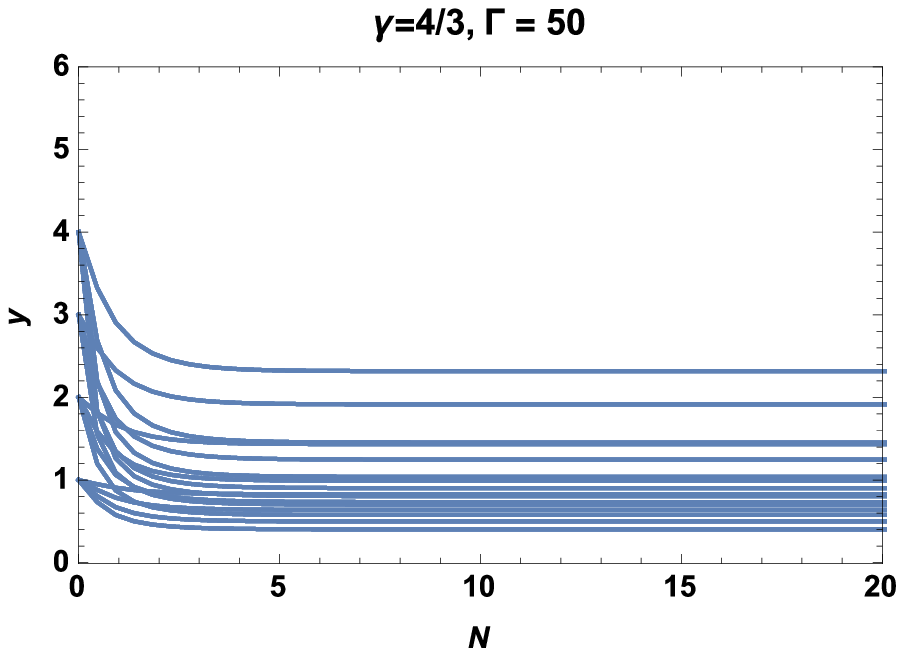}%
}
\subfloat[Projection of perturbations along $\lambda$ axis.]{%
  \includegraphics[clip,width=0.65\columnwidth]{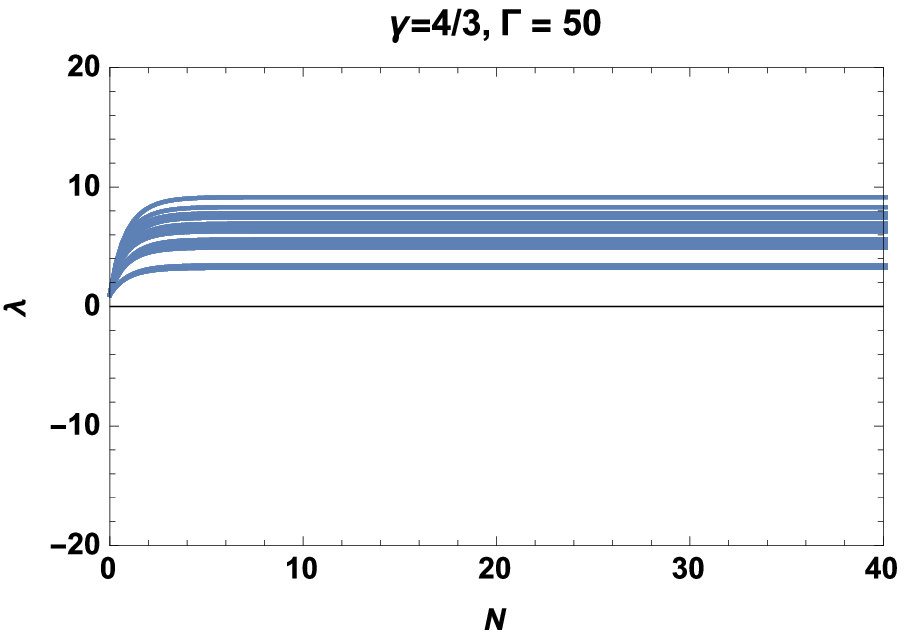}%
}

\caption{Projection of perturbations along $x,y , \lambda$ axis for Class II, without potential and $\gamma = \frac{4}{3}$.}

\end{center}

\end{figure*}

\section{Analysis with potential}

For a more general analysis of the system we introduce a potential $V=V(\phi)$ in the action as,

\begin{widetext}
\begin{equation} \label{actionp}
\mathcal{S} = \frac{1}{16 \pi} \int d^4 x \sqrt{- g}[ \phi R - \frac{\omega(\phi)}{\phi} g^{ab} \nabla_a \phi \nabla_b \phi - V(\phi)] + \mathcal{S}^{(m)}.
\end{equation} 

\end{widetext}

 The field equations and the wave equation are written as follows \\

\begin{widetext}
\begin{equation} \label{frid1p}
	3 H^2 = \frac{8 \pi}{\phi} \rho_{m} + \frac{\omega}{2} \frac{\dot{\phi}^2}{\phi^2} - 3 H \frac{\dot{\phi}}{\phi} + \frac{V}{2 \phi},
\end{equation}

\begin{equation} \label{frid2p}
\dot{H} = - \frac{\omega}{2} (\frac{\dot{\phi}}{\phi})^2 - \frac{8 \pi}{\phi} \rho_{m} (\frac{2 + \gamma \omega}{3 + 2 \omega}) + 2 H \frac{\dot{\phi}}{\phi} + \frac{1}{2 (2 \omega + 3) \phi} [\frac{d \omega}{d \phi} \dot{\phi}^2 - 2V + \phi \frac{dV}{d \phi}],
\end{equation}

\begin{equation} \label{wavep}
\ddot{\phi} + 3 H \dot{\phi} = \frac{1}{2 \omega + 3}[ 8 \pi (4 - 3 \gamma) \rho_{m}  - \phi \frac{dV}{d \phi} + 2V]- \frac{\dot{\phi}^2}{2 \omega + 3} \frac{d \omega}{d\phi}.
\end{equation}
\end{widetext}

The new set of variables has only one additional quantity $z$,

 $x = \frac{\dot{\phi}}{H \phi}, y = \frac{1}{2 \omega + 3}, z^2 = \frac{V}{6 H^2 \phi}, \lambda = \phi \frac{d \omega}{d \phi}$.

In what follows, we consider a power-law potential $V(\phi) = M^4 (\frac{\phi}{\phi_0})^{2n}$. The system is written down for a general $n$, but two definite examples, namely $n=1$ and $n=2$, will be worked out in detail. The choice is indeed motivated by simplicity of the formulation, but power law potentials are relevant as well. For  a very brief summary of why power law potentials are useful, we refer to a recent work\cite{soumya}. With the transformation of variables, the system of equations reduces to

\begin{widetext}

\begin{equation} \label{xv}
x^{\prime} = -\frac{3}{2} (1 + x - \frac{(1 - 3y)}{12 y} x^2  - z^2) [(2 - \gamma) x -  (4 - 3 \gamma) (2 + x)y] - \lambda x^2 y - \frac{1}{2} \lambda x^3 y - 3 x z^2 - 2 (6 n -1) y z^2 - 6(n - 1) x y z ^2,
\end{equation}

\begin{equation} \label{yv}
y^{\prime} = - 2 \lambda x y^2,
\end{equation}

\begin{equation} \label{zv}
z^\prime = (n - 1/2) x z - z[- \frac{3}{2} (( 4 - 3 \gamma) y + \gamma ) [1 + x - \frac{(1 - 3y)}{12 y} x^2  - z^2] - \frac{(1 - 3 y)}{4 y} x^2 + 2 x + \frac{1}{2} \lambda x^2 y - 6 (n - 1) y z^2]
\end{equation}
\begin{equation} \label{lambdav}
\lambda^{\prime} = ({\Gamma}+ {\lambda}) x,
\end{equation}

\end{widetext}

where $N = \ln(\frac{a}{a_{0}})$.

Similar to the previous case, here also we classify the system into two classes, I) ${\Gamma}+ {\lambda} = 0$, II) ${\Gamma}+ {\lambda} \neq 0$.  Each of these classes are also classified into two more sub classes $\gamma \neq 4/3$ and $\gamma = 4/3$.

\subsection{Class I: ${\Gamma} + \lambda = 0$}

In this case $\lambda$ is a constant then we have an effectively 3 dimensional system.
\subsection*{$i) ~\gamma \neq 4/3$}   
   
   The fixed points and the corresponding eigenvalues of the system are given in the table I,
   \begin{table}[h] 
\caption{}
\begin{center}
   \begin{tabular}{|c|c|c|c|c|}
   \hline 
  \makecell{ Fixed \\ points} & x & y & z & Eigenvalues \\ 
   \hline 
   a & 0 & $y \rightarrow 0$ & 0 & $\frac{3}{2} (\gamma - 2), 0,\frac{3}{2} \gamma, $ \\ 
   \hline 
   b & 0 & $y \rightarrow 0$ & 1 & $-3, 0, -3 \gamma, $ \\ 
   \hline 
   c & 0 & $y \rightarrow 0$ & -1 & $-3, 0, -3 \gamma, $ \\ 
   \hline 
   d & 0 & $\frac{\gamma}{6 \gamma - 8}$ & $-\sqrt{\frac{4 - 3 \gamma}{4 n - 3 \gamma}}$ & \makecell{ For $n = 1$ \\ ($-3,-3,0$)  \\  For n = 2 \\ ($ \frac{1}{5} (-3 \pm \sqrt{201}),0$)  }\\ 
   \hline 
   e & 0 & $\frac{\gamma}{6 \gamma - 8}$ & $\sqrt{\frac{4 - 3 \gamma}{4 n - 3 \gamma}}$ & \makecell{ For $n = 1$ \\ ($-3,-3,0$)  \\  For n = 2 \\ ($ \frac{1}{5} (-3 \pm \sqrt{201}),0$)  }\ \\ 
   \hline 
   \end{tabular} 
  \end{center}
  \end{table}

   The eigenvalues of the fixed points (a),(b) and (c) are independent of $n$ but the eigenvalues of (d) and (e) do depend on $n$. As mentioned before, we analyze the system for $n = 1$ and $n=2$.  All of these fixed points are normally hyperbolic. For $\gamma > 0$, fixed point (a) is a saddle point, while (b) and (c) are stable fixed points . Stability of the fixed points (d) and (e) are different for various choice of n, stable for $n=1$ and unstable for $n = 2$. To support our analytical  finding we have perturbed the system from the saddle point (a) and allowed it to evolve. In Fig 4 the plots are given and one can see that the system evolve to the fixed point (b) from (a). The plots are for $n=1$. Unlike $n=2$, the solutions for $n=1$ has attractors (d) and (e)  where $y \rightarrow  0$ is not a requirement.  But $n=2$ can only evolve to (b) and (c) and both of them implies $\omega \rightarrow \infty$.
   
   \begin{figure}[h]
\begin{center}

\subfloat[Evoluation of $x$.]{%
  \includegraphics[clip,width=0.6\columnwidth]{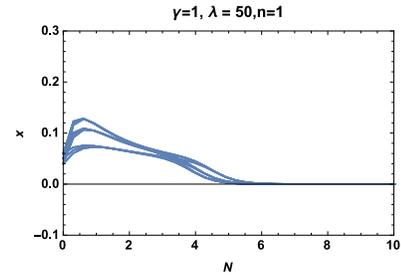}%
}

\subfloat[Evoluation of $y$.]{%
  \includegraphics[clip,width=0.6\columnwidth]{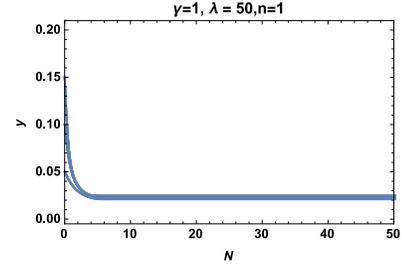}%
}

\subfloat[Evoluation of $\lambda$.]{%
  \includegraphics[clip,width=0.6\columnwidth]{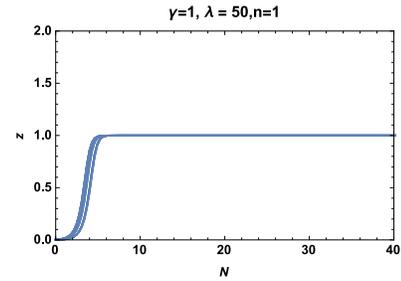}%
}

\caption{Evaluation of the system from the fixed point (a) to the fixed point (b) for Class I, with potential $V(\phi) = M^4 (\frac{\phi}{\phi_0})^{2n}, n=1$ and $\gamma \neq \frac{4}{3}$.}

\end{center}

\end{figure}

 \subsection*{$ii) ~\gamma = 4/3$}   
   
   The fixed points of the system are given in the Table II. These fixed points are also normally hyperbolic.
   
   \begin{table}[h] 
\caption{}
\begin{center}
\begin{tabular}{|c|c|c|c|c|}
   \hline 
   Fixed pointa & x & y & z & Eigenvalues \\ 
   \hline 
   a & 0 & $y$ & 0 & $(-1,0,2)$ \\ 
   \hline 
   b & 0 & $y \rightarrow 0$ & 1 & $(-3,0,-4)$ \\ 
   \hline 
   c & 0 & $y \rightarrow 0$ & -1 & $(-3,0,-4)$ \\ 
   \hline 
   \end{tabular}    
    \end{center}
  \end{table}
   
The value of $y$ in fixed point (a) is undefined and it is a saddle point. Fixed point (b) and (c) are stable fixed points.  The evolution of the system from some arbitrary initial values to (b) is shown in Fig 5.

  \begin{figure}[h]
\begin{center}

\subfloat[Evolution of $x$.]{%
  \includegraphics[clip,width=0.6\columnwidth]{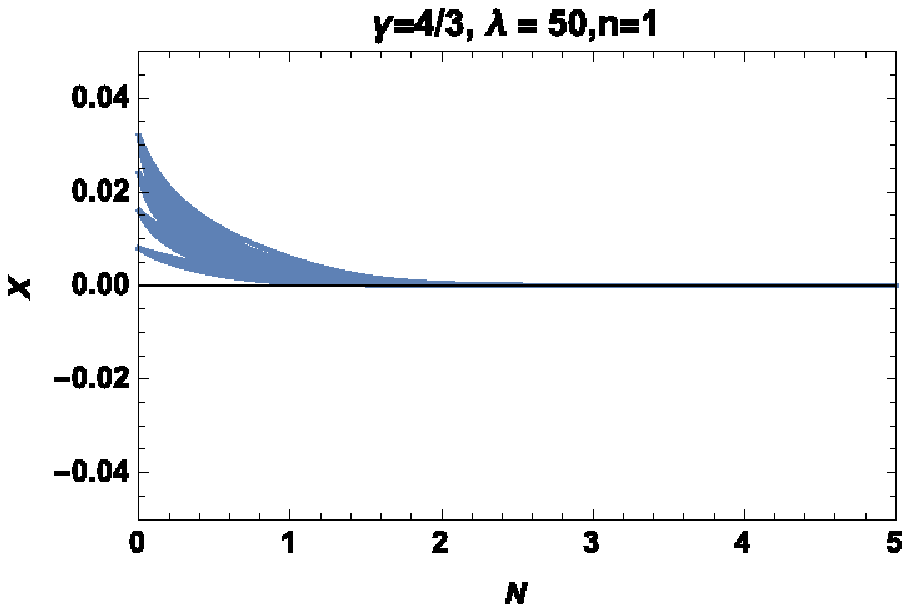}%
}

\subfloat[Evolution of $y$.]{%
  \includegraphics[clip,width=0.6\columnwidth]{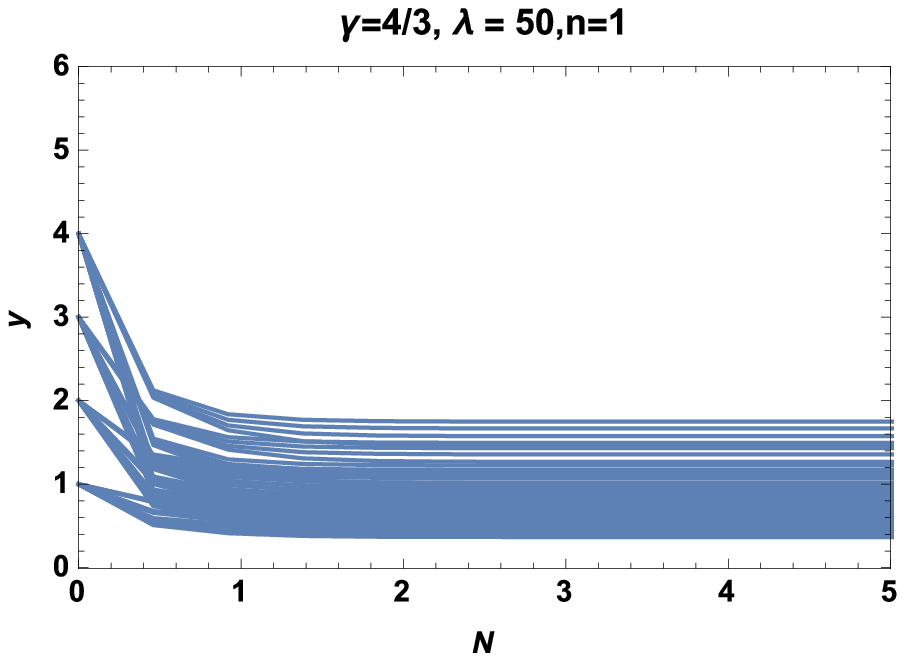}%
}

\subfloat[Evolution of $\lambda$.]{%
  \includegraphics[clip,width=0.6\columnwidth]{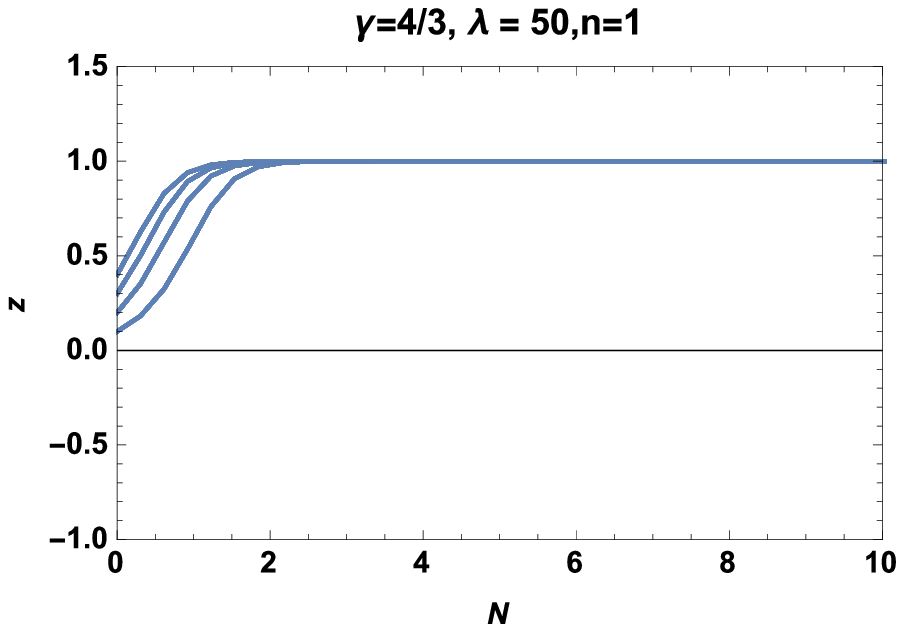}%
}

\caption{Evaluation of the system from some arbitrary initial values to the fixed point (b) for Class I, with potential $V(\phi) = M^4 (\frac{\phi}{\phi_0})^{2n}, n=1$ and $\gamma = \frac{4}{3}$.}

\end{center}

\end{figure}
   We have the plots for $n=1$. For $n =2$, the corresponding qualitative behaviour of Fig 5 are the same.
  \section{Class II: ${\Gamma} + \lambda \neq 0$}

  In this case the system is a 4-dimensional one. Like the previous scenario without a potential, here also we shall subdivide the class into two parts, one with $\gamma \neq \frac{4}{3}$ and the other with $\gamma = 4/3$. 

    \subsection*{$i)~ \gamma \neq 4/3$}   
    the fixed points and their corresponding eigenvalues are given in Table III. 

   \begin{table}[H] 
\caption{}
\begin{center}
    \begin{tabular}{|c|c|c|c|c|c|}
   \hline 
  \makecell{ Fixed \\ points} & x & y & z &$\lambda$& Eigenvalues \\ 
   \hline 
   a & 0 & $y \rightarrow 0 $ & 0 & $\lambda$ & $\frac{3}{2} (\gamma - 2), 0,\frac{3}{2} \gamma, 0$ \\ 
   \hline 
   b & 0 & $y \rightarrow 0$ & 1  & $\lambda$ & $-3, 0, -3 \gamma, 0$ \\ 
   \hline 
   c & 0 & $y \rightarrow 0$ & -1  & $\lambda$ & $-3, 0, -3 \gamma, 0$ \\ 
   \hline 
   d & 0 & $\frac{\gamma}{6 \gamma - 8}$ & $-\sqrt{\frac{4 - 3 \gamma}{4 n - 3 \gamma}}$  & $\lambda$ & \makecell{ For $n = 1$ \\ ($-3,-3,0,0$)  \\  For n = 2 \\ ($ \frac{1}{5} (-3 \pm \sqrt{201}),0,0$)  }\\ 
   \hline 
   e & 0 & $\frac{\gamma}{6 \gamma - 8}$ & $\sqrt{\frac{4 - 3 \gamma}{4 n - 3 \gamma}}$  & $\lambda$ & \makecell{ For $n = 1$ \\ ($-3,-3,0,0$)  \\  For n = 2 \\ ($ \frac{1}{5} (-3 \pm \sqrt{201}),0,0$)  }\ \\ 
   \hline 
   \end{tabular} 
   \end{center}
  \end{table}
  
  All these fixed pints are non hyperbolic and $\lambda$ is arbitrary for each fixed point.  We can not use linear stability analysis for this set of fixed points. However, for $2>\gamma > 0$, fixed point (a) is always a saddle point. For $n =2$, the fixed points (b) and (c) have the possibility of being late time attractors. These two fixed points are particularly interesting, as these two fixed points correspond to General Relativity.  We perturbed the system from the fixed points (b) and (c) and find that the system comes back to these fixed points so we conclude these fixed points to be stable.  Our system of equations is symmetric in $z \rightarrow -z$ and fixed point (c) has the same eigenvalues of (b) so the perturbations around (c) behave the same way as in (b). Fig 6 shows the projection of perturbations along the axes near the fixed point (b). 
  
   \begin{figure}[h]
\begin{center}

\subfloat[Projection of perturbations along $x$ axis.]{%
  \includegraphics[clip,width=0.6\columnwidth]{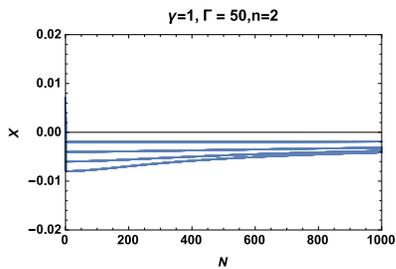}%
}

\subfloat[Projection of perturbations along $y$ axis.]{%
  \includegraphics[clip,width=0.6\columnwidth]{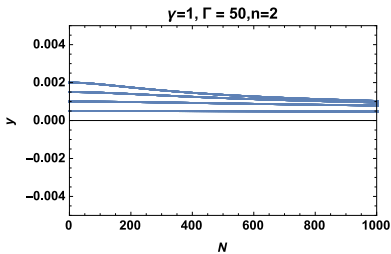}%
}

\subfloat[Projection of perturbations along $z$ axis.]{%
  \includegraphics[clip,width=0.6\columnwidth]{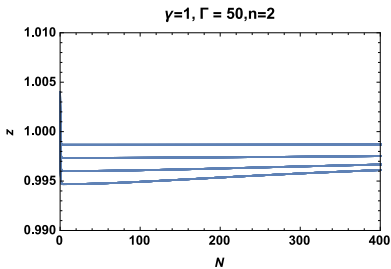}%
}

\subfloat[Projection of perturbations along $\lambda$ axis.]{%
  \includegraphics[clip,width=0.6\columnwidth]{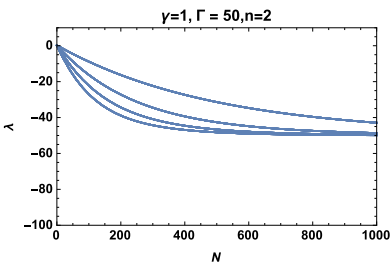}%
}
\caption{Projection of perturbations along $x,y ,z, \lambda$ axis for Class II, with potential $V(\phi) = M^4 (\frac{\phi}{\phi_0})^{2n}, n=2$ and $\gamma \neq \frac{4}{3}$.}

\end{center}
\end{figure}

 For $n=1$ there are more options because the fixed point (b), (c), (d) and (e) may be also late time attractor. To check the stability of (b) and (c), we allowed the system to evolve from some arbitrary initial values. Fig (7) confirms that the solutions are essentially attracted towards (b), hence the BD theory for $n=1$, in late time, has the possibility to be indistinguishable  from general relativity.

   \subsection*{ $ii) ~ \gamma = 4/3$}   
   
      \begin{table}[H] 
\caption{}
\begin{center}
   \begin{tabular}{|c|c|c|c|c|c|}
   \hline 
   Fixed pointa & x & y & z & $\lambda$ & Eigenvalues \\ 
   \hline 
   a & 0 & $y \rightarrow 0$ & 0 & $\lambda$& $(-1,0,2,0)$ \\ 
   \hline 
   b & 0 & $y \rightarrow 0$ & 1& $\lambda$ & $(-3,0,-4,0)$ \\ 
   \hline 
   c & 0 & $y \rightarrow 0$ & -1 & $\lambda$& $(-3,0,-4,0)$ \\ 
   \hline 
   \end{tabular}    
   \end{center}
  \end{table}
  
The fixed points and the eigenvalues are given in Table IV. There are only three fixed points and all of them are nonhyperbolic. Fixed point (a) is a saddle fixed point  but (b) and (c) may be late time attractors. Similar to previous cases we find the stability of these fixed point (b) and (c) numerically. Fig 8, shows the evolution of the perturbations around the fixed point (b). The phase space behaviour of (b) and (c) are similar.  Hence, we conclude (b) and (c) are indeed stable fixed points. \\

The major difference in the stability in this cases, where a potential $V(\phi)$ is also included in the action, is the fact that the requirement of an infinite $\omega$ for matter distribution with a non-zero trace is relaxed in some cases. On the other hand, for a trace-free matter like a radiation, an infinite $\omega$ seems to be a requirement as opposed to the purer version of the theory, i.e., in the absence of the potential term. The reason seems to be quite apparent. The presence of the potential term, $-\phi \frac{dV}{d\phi} + 2V$, in the equation (\ref{wavep}) makes the $\phi$ dependence of the solution for $\omega$ quite open. The potential term can conspire with the $\frac{d\omega}{d\phi}$ term and infringe upon the $\omega$ dependence of the scalar field $\phi$.

 \begin{figure}[h]
\begin{center}

\subfloat[Evolution of $x$ .]{%
  \includegraphics[clip,width=0.6\columnwidth]{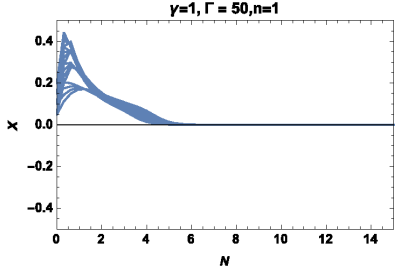}%
}

\subfloat[ Evolution of $y$.]{%
  \includegraphics[clip,width=0.6\columnwidth]{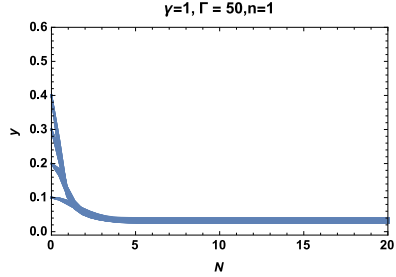}%
}

\subfloat[Evolution of $z$.]{%
  \includegraphics[clip,width=0.6\columnwidth]{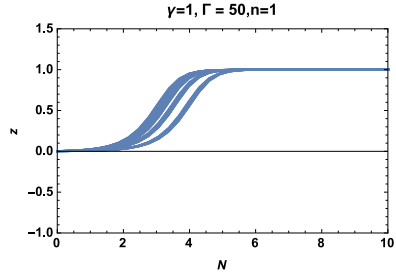}%
}

\subfloat[Evolution of $\lambda$.]{%
  \includegraphics[clip,width=0.6\columnwidth]{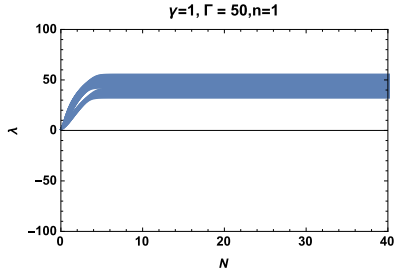}%
}
\caption{Evolution of the system from some arbitrary initial conditions to the fixed point(b) for Class II, with potential $V(\phi) = M^4 (\frac{\phi}{\phi_0})^{2n}, n=1$ and $\gamma \neq \frac{4}{3}$.}
\end{center}
\end{figure}

\begin{figure}[H]
\begin{center}
\subfloat[Projection of perturbations along $x$ axis.]{%
  \includegraphics[clip,width=0.6\columnwidth]{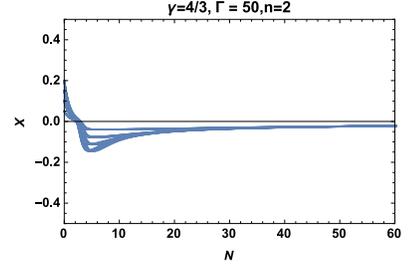}%
}

\subfloat[Projection of perturbations along $y$ axis.]{%
  \includegraphics[clip,width=0.6\columnwidth]{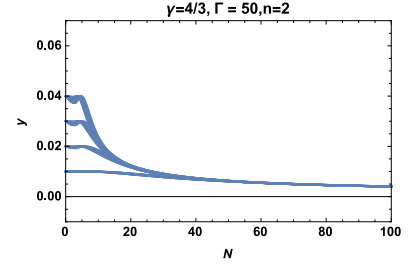}%
}

\subfloat[Projection of perturbations along $z$ axis.]{%
  \includegraphics[clip,width=0.6\columnwidth]{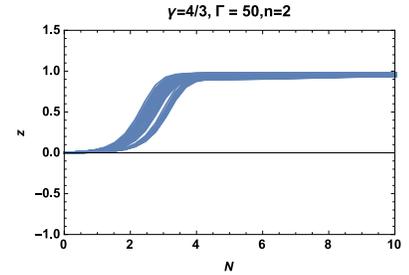}%
}

\subfloat[Projection of perturbations along $\lambda$ axis.]{%
  \includegraphics[clip,width=0.6\columnwidth]{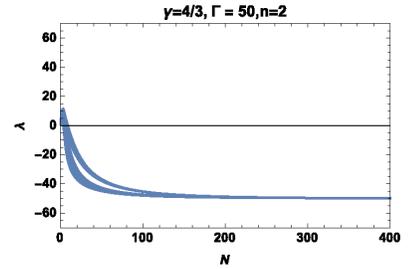}%
}
\caption{Projection of perturbations along $x,y , z, \lambda$ axis for Class II, with potential $V(\phi) = M^4 (\frac{\phi}{\phi_0})^{2n}, n=2$ and $ \gamma = \frac{4}{3}$.}

\end{center}
\end{figure}

\section{Discussions}

The present work deals with the stability of cosmological solutions for a spatially flat homogeneous and isotropic cosmological model with a distribution of a perfect fluid in an extended Brans-Dicke theory. The extension actually lies in the generalization of the parameter $\omega$ as a function of the BD field $\phi$ according to the prescription by Nordtvedt\cite{nord}. The strength of Nordtvedt's theory lies in the fact that this can reproduce a wide class of scalar-tensor theory of gravity for suitable choices of $\omega = \omega (\phi)$. \\

For a particular choice of $\omega$, given by $\omega (\phi) = \ln (B \phi^A)$, the problem actually reduces 
to a 2-dimensional one. It is found that for any fluid with a reasonable equation of state other than radiation, the solution is stable only for $\phi=$ constant and $\omega \rightarrow \infty$, for which the theory is indistinguishable from general relativity! However, for a radiation distribution, the situation different, a constant $\phi$ is still a requirement for the stability but $\omega$ does not need to be infinity, it can have any value, as $y$, which is $\frac{1}{2\omega + 3}$, can have an arbitrary value so as to allow a stable situation. \\

Even for all other functional forms of $\omega$, the criteria of stability are very similar. When $\gamma \neq \frac{4}{3}$, there is only one fixed point. This fixed point is a nonhyperbolic fixed point, and the stability cannot be judged analytically. A numerical perturbation about the fixed point clearly indicates that $\phi=$ constant and $\omega \rightarrow \infty$ is indeed a stable natural habitat for the system.  However, for $\gamma = \frac{4}{3}$ corresponding to radiation, where again due to the nonhyperbolic nature of the fixed point, only the stability against perturbation can be analysed. Any value of $\omega$ can potentially give rise to a stable situation (see figure 3). However, $\phi$ certainly approaches a constant value for the stable solutions ($x=0$). A constant $\phi$ indicates that the scalar field is trivial and makes no impact on the geometry.\\

One should note that for a radiation distribution ($\gamma = \frac{4}{3}$), the trace of energy momentum tensor is zero, whereas for other values of $\gamma$,  the trace is nonzero. Now we find an intriguing result that the stability of the generalized BD solutions warrants an infinite value of $\omega$ for matter fields with a nonzero trace for the energy momentum tensor, whereas for a trace free matter distribution, an infinite $\omega$ has nothing to do with the stability. This result reminds the fact that an infinite $\omega$ leads BD theory to the corresponding GR for matter with a nonzero trace and BD theory with a traceless matter does not have this infinite $\omega$ limit to GR\cite{soma, valerio1}. So although nothing definite can be talked about a radiation distribution, for any other physically relevant fluid distribution, the natural habitat for the BD theory for a large $N$, i.e., a far future, is indeed general relativity. \\

A further generalization of BD theory, namely the inclusion of a scalar potential $V(\phi)$ in the Lagrangian, is also considered in section III. This generalization renders the theory significantly different from BD theory, but it has a relevance, particularly in connection with the building up of models for the accelerating universe (see for example, reference \cite{bertolami}). For the sake of simplicity, we have taken up the case for a power law potential ($V(\phi) = M^4 (\frac{\phi}{\phi_0})^{2n}$) and work out in detail for $n=1$ and $n=2$. Although specific examples do not have the same status as a general treatment, but there are clear indications that with the addition of a potential, the requirement of infinite $\omega$ for a stability is relaxed in some of the cases where the trace of the matter energy momentum tensor ($T$) is non zero. Surprisingly, this infinite $\omega$ limit becomes more relevant where $T=0$. For the latter the explanation could well be that in the presence of $V(\phi)$, the matter Lagrangian now has trace, and the stability requirement becomes similar to $T \neq 0)$ case of the theory without a potential. \\

As a matter of clarification, we should mention what is meant by the statement BD theory reduces to GR or not. This is particularly important as in all the fixed points one has $x=0$ meaning a constant $\phi$. It is easy to see that for a constant $\phi$, the scalar field contribution to the field equation (\ref{field}) becomes trivial and it reduces practically some equation in GR. But it might lead to some other matter distribution than that one starts with. When one says that BD theory reduces to GR, it is meant that the corresponding equation in GR with exactly the same matter distribution\cite{soma}.

 \end{document}